\documentclass[aps,prd,twocolumn,nofootinbib]{revtex4-1}
\makeatletter

\newcommand{\Rmnum}[1]{\expandafter\@slowromancap\romannumeral #1@}
\makeatother
\usepackage[colorlinks,linkcolor=blue,anchorcolor=blue,citecolor=blue,urlcolor=blue,breaklinks=true]{hyperref}

\usepackage{epsfig}
\usepackage{natbib}
\usepackage{epstopdf}
\usepackage{mathrsfs}
\usepackage{slashed}
\usepackage{amsmath}
\usepackage{verbatim}
\usepackage{graphicx}
\usepackage{amssymb}
\usepackage{psfrag}
\usepackage{array}
\newcommand{\ud}{\mathrm{d}}
\newcommand{\ue}{\mathrm{e}}
\newcommand{\ui}{\mathrm{i}}

\newcommand{\bp}{\boldsymbol{p}}

\newcommand{\MeV}{\mathrm{MeV}}

\begin{document}
\author{Yi-Lun Du$^{1}$}
\author{Ya Lu$^{1}$}
\author{Shu-Sheng Xu$^{1,3}$}
\author{Zhu-Fang Cui$^{1,3}$}
\author{Chao Shi$^{1,3}$}
\author{Hong-Shi Zong$^{1,2,3}$}\email{Email:zonghs@nju.edu.cn}

\address{$^{1}$ Department of Physics, Nanjing University, Nanjing 210093, China}
\address{$^{2}$ Joint Center for Particle, Nuclear Physics and Cosmology, Nanjing 210093, China}
\address{$^{3}$ State Key Laboratory of Theoretical Physics, Institute of Theoretical Physics, CAS, Beijing 100190, China}

\title{Susceptibilities and critical exponents within the Nambu--Jona-Lasinio model}
\begin{abstract}
In the mean field approximation of (2+1)-flavor Nambu--Jona-Lasinio model, we strictly derive several sets of coupled equations for the chiral susceptibility, the quark number susceptibility, etc. at finite temperature and quark chemical potential. The critical exponents of these susceptibilities in the vicinity of the QCD critical end point (CEP) are presented in SU(2) and SU(3) cases, respectively. It is found that these various susceptibilities share almost the same critical behavior near the CEP. The comparisons between the critical exponents for the order parameters and the theoretical predictions are also included.

\bigskip

\noindent PACS Number(s): 12.39.-x, 25.75.Nq, 12.39.Fe
\end{abstract}

\maketitle

\section{Introduction}
\label{sec:intro}
It is believed that the strongly-interacting matter undergoes a phase transition to the quark-gluon plasma (QGP) at high temperature $T$ and/or at high quark chemical potential $\mu$. Partially restored chiral symmetry and deconfined quarks and gluons are expected in this new state of matter~\cite{Masayuki1989668,klimt1990chiral,ratti2006phases,roessner2008chiral,PhysRevD.77.114028,fu20082+}. The phase transition should have implications for the study of the on-going heavy-ion collision experiments at the BNL Relativistic Heavy-Ion Collision (RHIC) and the Large Hadron Collider (LHC)\cite{Heinz:2004qz,jacobs2005matter,wang2014hard,Antinori:2014xma} and the future FAIR(GSI) project in Darmstadt and NICA project in Dubna~\cite{1742-6596-50-1-048,friese2007cbm,0954-3899-36-6-064069}.

The phase transition from the hadronic matter to the QGP at finite $T$ with vanishing $\mu$ has been  investigated extensively on the lattice and within the effective theoretical models. It is believed that with two massless quarks the chiral phase transition is of second order and the critical behavior falls into the universality class of the $O(4)$ Heisenberg magnet spin model in three dimensions~\cite{wilczek1992application,Rajagopal1993395}. However, in nature, the u- and d-quark have small but finite masses and the second order phase transition turns into a smooth crossover.

The reliable lattice simulations of the QCD chiral transition with finite $\mu$ are not available due to the severe fermion sign problem. Many effective theories~\cite{Masayuki1989668,klevansky1992nambu,alford1998qcd,PhysRevD.58.096007,costa2008thermodynamics} of QCD predict the existence of the tricritical point (TCP) and CEP in the QCD $T-\mu$ phase diagram in massless and massive quark cases, respectively. The TCP is the point where a line of critical points ($O(4)$ line, a second-order transition line) at lower $\mu$'s and higher $T$'s turns into a first-order phase transition line at higher $\mu$'s and lower $T$'s. With the current quark masses increasing from zero to the physical masses, the TCP emerges to the CEP at which the first-order phase transition line ends from higher $\mu$'s towards lower $\mu$'s in the phase diagram. The location of the CEP has been reported to be accessible at RHIC~\cite{PhysRevLett.113.092301}. 

Both at the TCP and CEP, second-order phase transitions are considered to occur, which are characterized by the long-wavelength fluctuations of the order parameter. Therefore, the values of critical exponents will be an interesting and important question, which is related to the universality class of this phase transition. Based on the analysis of the Landau-Ginzburg thermodynamic potential's expansion in the order parameter, the critical exponents in mean field approximation have been predicted~\cite{PhysRevD.67.014028}. Furthermore it's expected that the critical exponents of the susceptibility depends on whether the path to approach the CEP is asymptotically parallel to the first order transition line or not~\cite{PhysRevA.2.1047}. It is necessary to verify these universality arguments from the calculations of specific effective theory of QCD~\cite{schaefer2007susceptibilities}. The Nambu--Jona-Lasinio (NJL) model is believed to be a good choice~\cite{costa2007qcd,sasaki2008chiral,costa2008thermodynamics,costa2009qcd,costa2010phase}.

The rest of this paper is organized as follows. In Sec. \Rmnum{2} the formulation of the (2+1)-flavor NJL model and the QCD thermodynamics are briefly reviewed. In Sec. \Rmnum{3} we present the derivation of several sets of coupled equations for chiral susceptibility, quark number susceptibility, etc. at finite temperature and quark chemical potential. In Sec. \Rmnum{4}, the numerical results of several susceptibilities and their  corresponding critical exponents are presented. It will be found that they are consistent with the predictions by the formalism of Landau-Ginzburg thermodynamic potential's expansion in the order parameter~\cite{PhysRevD.67.014028}. The discussions on the critical exponents for the order parameters at the CEP and vanishing chemical potential in the chiral limit are also included. Finally, in Sec.\Rmnum{5} we will summarize our results and give the conclusions.

\section{The Nambu--Jona-Lasinio model}
The Lagrangian of the SU(3) NJL model~\cite{PhysRev.122.345,PhysRev.124.246,klevansky1992nambu,buballa2005njl,hatsuda1994qcd} is given by
\begin{equation}
\begin{aligned}
\mathscr{L} =&\bar{\psi}({\ui}\slashed{\partial}-\hat{m})\psi+G_1[(\bar{\psi}\lambda_i\psi)^2+(\bar{\psi}{\ui}\gamma_5\lambda_i\psi)^2]\\
            &-G_2[(\bar{\psi}\lambda_i\gamma_\mu\psi)^2+(\bar{\psi}\lambda_i\gamma_\mu\gamma_5\psi)^2]\\
            &-K\{{\det}[\bar{\psi_i}(1+\gamma_5)\psi_j]+{\det}[\bar{\psi_i}(1-\gamma_5)\psi_j]\},
\end{aligned}
\end{equation}
where the column vector $\psi=(u,d,s)$ is the quark field with three flavors, $N_f=3$, and three colors, $N_c=3$. $\hat{m}=\mathrm{diag}(m_u,m_d,m_s)$ is the current quark mass matrix. $\lambda^a$ are the Gell-Mann matrices, a=1,2,3,...,8, $\lambda^0=\sqrt{\frac{2}{3}}\mathbf{I}$. The second and third terms represent four-point scalar-pseudoscalar (SP) and vector-axialvector (VA) interactions, respectively, which are invariant under the chiral $\mathrm{U(3)_L}\otimes\mathrm{U(3)_R}$ symmetry. The last term represents six-point interaction which breaks the axial $\mathrm{U(1)_A}$ symmetry, but it is invariant under $\mathrm{SU(3)_L}\otimes\mathrm{SU(3)_R}$. These four-point and six-point interactions are constructed in accordance with all relevant symmetries of QCD.

The color singlet part of the six-point interaction can be written
\begin{equation}
\begin{aligned}
\mathscr{L}_6=&\frac{K}{6}D_{ijk}[\frac{1}{3}(\bar{\psi}\lambda_i\psi)(\bar{\psi}\lambda_j\psi)(\bar{\psi}\lambda_k\psi)\\
&-(\bar{\psi}{\ui}\gamma_5\lambda_i\psi)(\bar{\psi}{\ui}\gamma_5\lambda_j\psi)(\bar{\psi}\lambda_k\psi)].
\end{aligned}
\end{equation}
The totally symmetric coefficients $D_{ijk}$ are identical to the SU(3) structure constants $d_{ijk}$ for $i,j,k\in\{1,2,3,...,8\}$; $D_{0jk}=-1/\sqrt{6}$ for $j=k=1,2,3,...,8$ and $D_{000}=\sqrt{\frac{2}{3}}$.

Developing the thermodynamics of $\mathscr{L_{\text{NJL}}}$ in the mean field approximation, we obtain the thermodynamical potential density $\omega=\Omega/V$. The conditions $\partial\omega/\partial(\delta m)=\partial\omega/\partial(\delta \mu)=0$, which minimize the thermodynamical potential, lead to the gap equations for each of the quark flavors $i$:

\begin{equation}
\delta m_i\equiv M_i-m_i=-4G_1\langle\!\langle\bar{q_i}q_i\rangle\!\rangle+2K\langle\!\langle\bar{q_j}q_j\rangle\!\rangle\langle\!\langle\bar{q_k}q_k\rangle\!\rangle,\label{eqn:M}
\end{equation}
with $i, j, k$ cyclic, and
\begin{equation}
\delta \mu_i\equiv\mu_{ri}-\mu_i=4G_2\langle\!\langle q_i^\dagger q^{}_i\rangle\!\rangle.\label{eqn:mur}
\end{equation}
Where $M_i$ and $\mu_{ri}$ are the constituent quark mass and renormalized chemical potential of flavor $i$, respectively.
Here the thermal expectation value of an operator $\Theta$ is denoted as
\begin{equation}
\langle\!\langle\Theta\rangle\!\rangle=\frac{\mathrm{Tr}\,\Theta\,{\ue}^{-\beta(\mathscr{H}-\mu_i\mathscr{N}_i})}{\mathrm{Tr}\,
{\ue}^{-\beta(\mathscr{H}-\mu_i\mathscr{N}_i)}}.
\end{equation}
Where $\mathscr{H}$ is the Hamiltonian operator, while $\mathscr{N}$ is the quark number density operator. The chiral condensate $\langle\!\langle\bar{q}q\rangle\!\rangle$ and quark number density $\langle\!\langle q^\dagger q\rangle\!\rangle$ are given by
\begin{gather}
 \begin{split}
 \hspace{-5mm}\langle\!\langle\bar{q}q\rangle\!\rangle=-M\frac{N_c}{\pi^2}\!\int_0^\Lambda\!
  \frac{p^2}{E}\big\{1-n(\boldsymbol{p},\mu_r)-
  m(\boldsymbol{p},\mu_r)\big\}{\ud}p,\label{eq:sigma11}\\
  \end{split}\\
\begin{split}
 \hspace{-20mm} \langle\!\langle q^\dagger q\rangle\!\rangle=\frac{N_c}{\pi^2}\int_0^\Lambda
  p^2\big\{n(\boldsymbol{p},\mu_r)-m(\boldsymbol{p},\mu_r)\big\}{\ud}p, \label{eq:sigma22}
\end{split}
 \end{gather}
with
\begin{equation}\label{eqn:nm}
  n(\boldsymbol{p},\mu)=\frac{1}{1+\exp{[\beta(E-\mu)]}},
\end{equation}
\begin{equation}
  m(\boldsymbol{p},\mu)=\frac{1}{1+\exp{[\beta(E+\mu)]}},
\end{equation}
$E=\sqrt{M^2+{\bp}^2}$ and $\beta=1/T$. $\Lambda$ is a momentum cutoff, which is introduced to regulate the ultraviolet divergence. For convenience we will use the non-covariant three-momentum cutoff
scheme~\cite{PhysRevD.34.1601}.

The thermodynamical potential density $\omega=\Omega/V$ is given by
\begin{equation}\label{eqn:W}
\begin{aligned}
\omega=&\frac{-T}{V}\log [ \mathrm{Tr}\ \mathrm{e}^{-\beta(\mathscr{H}-\mu_i\mathscr{N}_i)}]\\
          =&\sum_{i=u,d,s}-\frac{N_c}{\pi^2}\int_0^\Lambda p^2 \{E_i+T(\log[1+\mathrm{e}^{\beta(\mu_{ri}-E_i)}]\\
          &+\log[1+\mathrm{e}^{\beta(-\mu_{ri}-E_i)}])\} dp+2G_1{\langle\!\langle\bar{q}_iq_i\rangle\!\rangle}^2\\
          &-2G_2{\langle\!\langle q^\dagger_i q^{}_i\rangle\!\rangle}^2-4K\langle\!\langle\bar{q}_uq_u\rangle\!\rangle\langle\!\langle\bar{q}_dq_d\rangle\!\rangle\langle\!\langle\bar{q}_sq_s\rangle\!\rangle.
\end{aligned}
\end{equation}

In this paper, the widely accepted parameter set will be employed~\cite{PhysRevC.53.410}: $m_{u,d}=5.5~{\MeV}$, $m_s=140.7~{\MeV}$, $\Lambda=602.3~{\MeV}$,
$G_1\Lambda^2=1.835$, $G_2=0$, $K\Lambda^5=12.36,$ which yields $f_\pi=92.4~{\MeV}$, $M_\pi=135.0~{\MeV}$, $M_K=497.7~{\MeV}$, $M_{\eta}=514.8~{\MeV}$, $M_{\eta^\prime}=957.8~{\MeV}$ and quark condensates ${\langle\!\langle\bar{q}_uq_u\rangle\!\rangle}^{1/3}={\langle\!\langle\bar{q}_dq_d\rangle\!\rangle}^{1/3}=-241.9~{\MeV}$, ${\langle\!\langle\bar{q}_sq_s\rangle\!\rangle}^{1/3}=-257.7~{\MeV}$. The constituent quark masses are also obtained: $M_u=M_d=367.7~{\MeV}$ and $M_s=549.5~{\MeV}$. Here we don't include the vector-axialvector (VA) interaction. However, for the generalization of the equations to calculate the susceptibilities shown below, we keep the $G_2$ coupling in our following derivation.

Now Eqs. (\ref{eqn:M}),(\ref{eqn:mur}),(\ref{eq:sigma11}) and (\ref{eq:sigma22}) form a set of self-consistent equations. By solving these equations iteratively, one can obtain the effective quark masses, chiral condensates and quark number densities at finite temperature and quark chemical potential (only the simple physical case $\mu_u=\mu_d=\mu_s$ is considered in this paper). In the co-existence region of Wigner solution (chiral symmetry partially restored) and Nambu solution (chiral symmetry broken), the solution which minimizes the thermodynamical potential density $\omega$ is stable~\cite{sasaki2008chiral}.
The thermodynamical potential density $\omega=\Omega/V$ is given in Eq. (\ref{eqn:W}).

\begin{figure}[tbp]
\begin{minipage}[t]{7cm}
\includegraphics[width=3in]{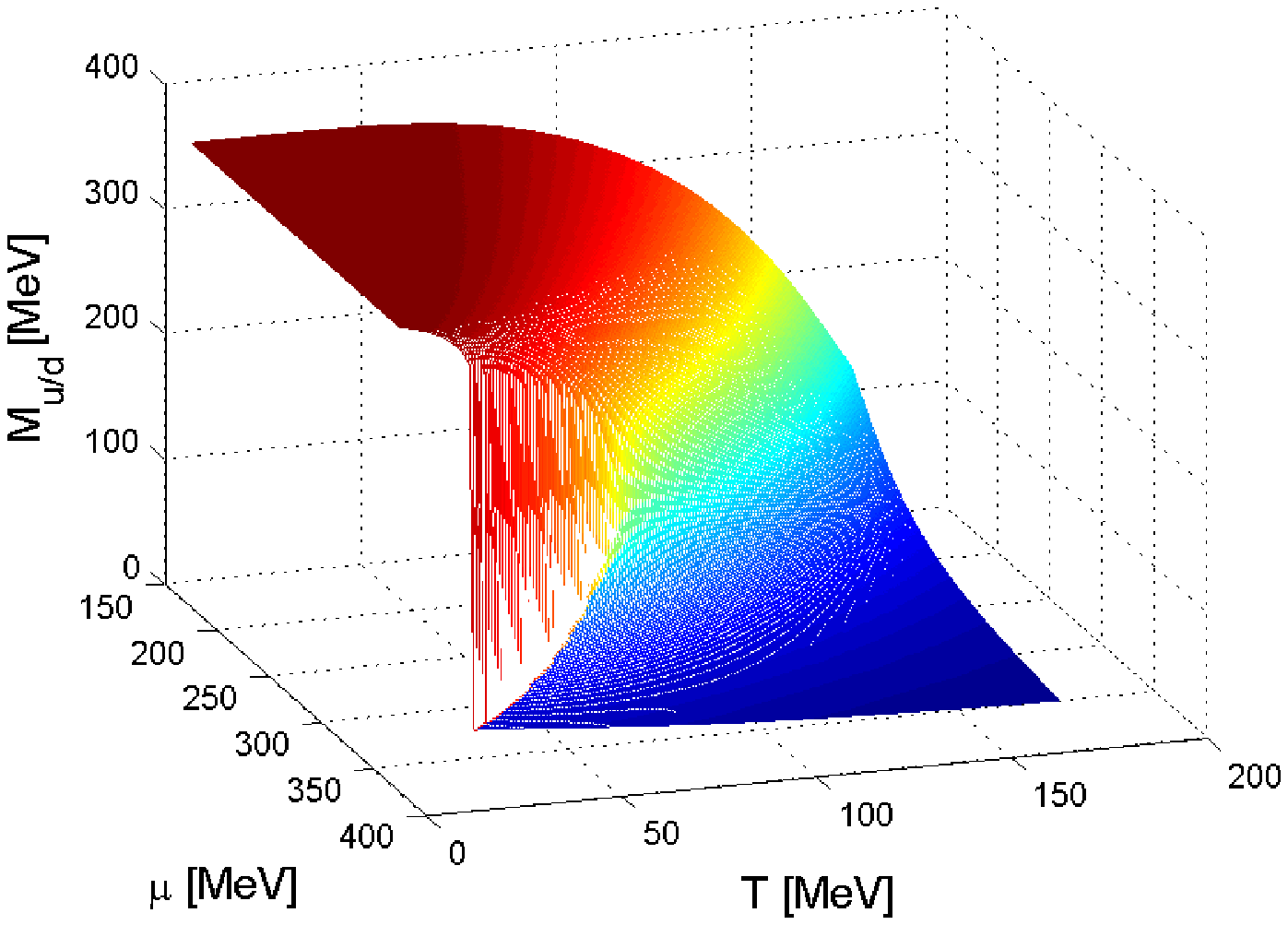}
\caption{(color online): Up/Down quark mass $M_{u/d}$ [MeV]}
\label{fig:M1}
\end{minipage}
\hfill
\begin{minipage}[t]{7cm}
\includegraphics[width=3in]{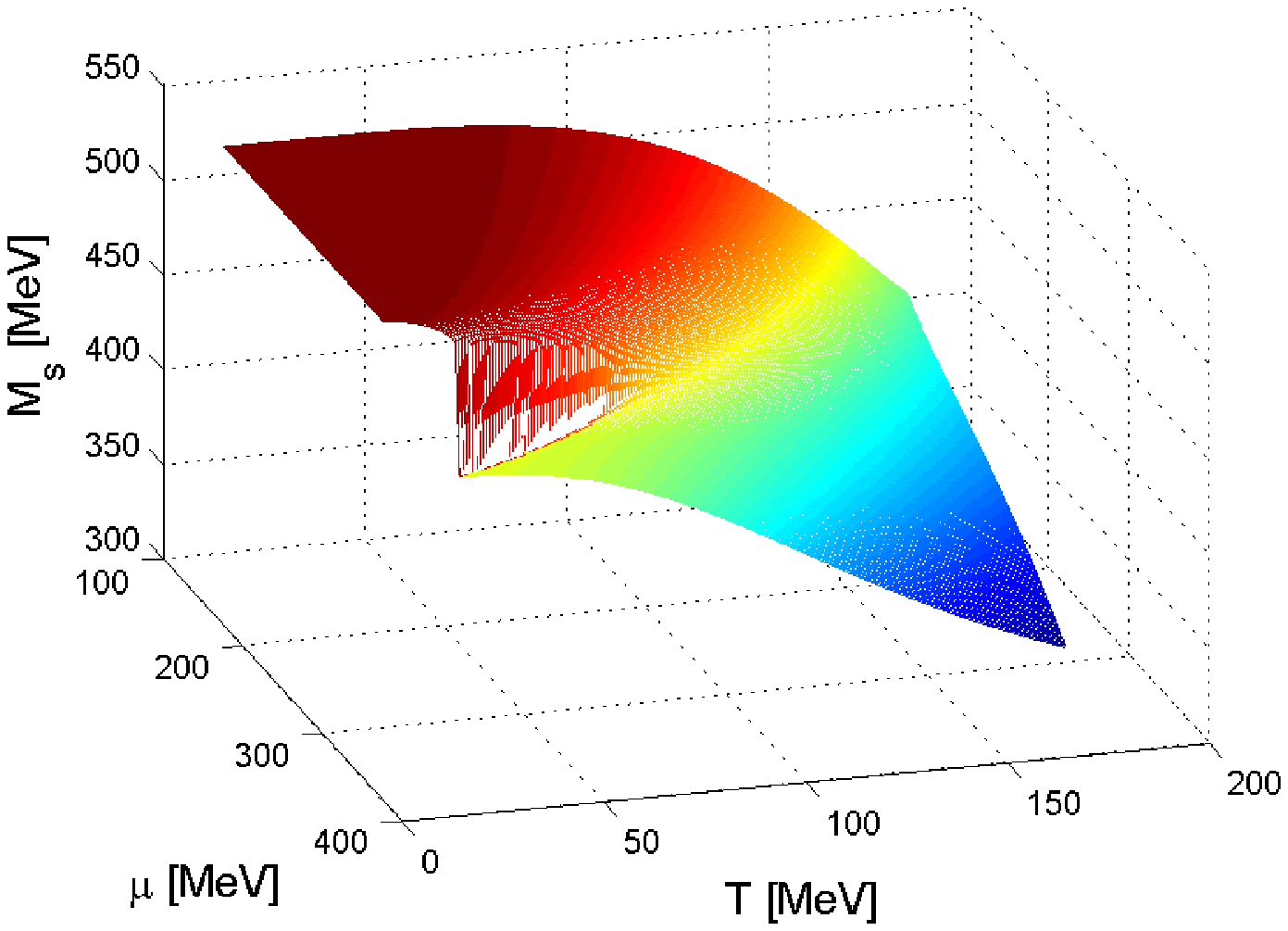}
\caption{(color online): Strange quark mass $M_s$ [MeV]}
\label{fig:Ms}
\end{minipage}
\end{figure}

Having taken this into account, we show the final results in Fig. \ref{fig:M1} and Fig. \ref{fig:Ms}. It's worth noting that, at $T$=0, $M_s$ stays almost constant after $\mu_{crit}$ where first-order transition occurs until $\mu$ exceeds $M_s$ and we start to have density of strange quarks in the system. However, at finite $T$, we start to have density of strange quarks at a relatively small $\mu$ in the system and this is a different feature compared with the $T$=0 case. According to our numerical calculation, the number density of s-quark is approximately one tenth of that of u(d)-quark near the CEP. Therefore we can expect that the presence of s-quark will influence the location of the CEP and the interplay of s-quark with u(d)-quarks has the potential to affect the critical behavior near the CEP. Hence we study critical behavior related with the strange sector in the vicinity of the CEP in the following.

\section{Exact equations for various susceptibilities}
Experimentally, the linear responses, such as the susceptibility, conductivity, etc., of the physical system to some external field, are often measured to study the properties of the related system. Therefore the studies of various susceptibilities, as linear responses of the quark matter to the external fields are very important on the theoretical side, which characterize the non-perturbative properties of QCD vacuum~\cite{PRC.72.035202,PLB.639.248--257,PLB.669.327--330,PRC.81.032201,AOP.358.172} and are also widely used to study the phase transitions of strongly interacting matter~\cite{ratti2007quark,JHEP.1407.014,PRD.90.036007,zhao2014nonlinear}.

Now let us introduce the definitions of six kinds of susceptibilities to be discussed in this work: the chiral susceptibility $\chi_s$, the quark number susceptibility $\chi_q$, the vector-scalar susceptibility $\chi_{vs}$, the thermal susceptibility $\chi_T$ and two auxiliary susceptibilities  $\chi_m$ and $\chi_n$. For mathematical convenience we first introduce four of these susceptibilities in the free quark gas case (the interaction terms in the Lagrangian are set to be zero, i. e., $\mathscr{L}_{int}=0$)~\cite{kunihiro2000chiral}, where $M$ and $\mu_r$ are reduced to $m$ and $\mu$, which are independent quantities. Denoted with the superscript $^{(0)}$, their definitions and expressions are as follows:
\begin{gather}
\begin{split}
 \hspace{-20mm}\chi_s^{(0)}\equiv&-\frac{\partial\langle\!\langle\bar{q}q\rangle\!\rangle_f}{\partial m}\\
  =&\frac{N_c}{\pi^2}\int_0^\Lambda\bigg[\frac{m^2p^2\beta}{E^2}g(\mu)+\frac{p^4}{E^3}f(\mu)\bigg]{\ud}p, \label{eq:chis}
\end{split}\\
\begin{split}
 \hspace{-14mm}\chi_q^{(0)}\equiv&\frac{\partial\langle\!\langle q^{\dagger}q\rangle\!\rangle_f}{\partial \mu}=\frac{N_c}{\pi^2}\int_0^\Lambda p^2\beta g(\mu){\ud}p, \label{eq:chiq}
\end{split}\\
\begin{split}
 \hspace{-12mm}\chi_{vs}^{(0)}\equiv&\frac{\partial\langle\!\langle\bar{q}q\rangle\!\rangle_f}{\partial \mu} =\frac{N_c}{\pi^2}\int_0^\Lambda\frac{mp^2\beta}{E}h(\mu){\ud}p, \label{eq:chivs}
\end{split}\\
\begin{split}
 \hspace{-32mm}\chi_{m}^{(0)}\equiv-\frac{\partial\langle\!\langle q^{\dagger}q\rangle\!\rangle_f}{\partial m}=\chi_{vs}^{(0)}, \label{eq:chim}
\end{split}
\end{gather}
where $g(\mu)+h(\mu)=2n(\mu)(1-n(\mu))$, $g(\mu)-h(\mu)=2m(\mu)(1-m(\mu))$, $f(\mu)=1-n(\mu)-m(\mu)$,
the subscript ``$f$" represents the free quark gas systems. It should be noted that $\chi_{m}^{(0)}$ and $\chi_{vs}^{(0)}$ have the same analytical expression, which is reasonable from the viewpoint of statistical mechanics:
\begin{equation}\label{eqn:mixed partial derivative}
\begin{aligned}
\chi_{m}^{(0)}=\chi_{vs}^{(0)}=\frac{T}{V}\frac{\partial^2}{\partial m \partial \mu}\ln Z_{f},
\end{aligned}
\end{equation}
where $Z_{f}$ is the QCD partition function in the free quark gas case.

\begin{figure}[tbp]
\centering
\includegraphics[width=3in]{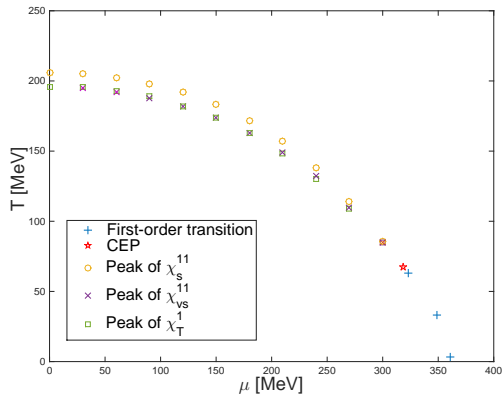}
\caption{(color online): Phase diagram with a critical band in the crossover region.}
\label{fig:phasedia}
\end{figure}

In the interacting case, $M_i$ and $\mu_{ri}$ ($i=u, d, s$) are no longer independent, they both are the functions of $m_1=m_u=m_d$, $m_2=m_s$, $\mu_1=\mu_u=\mu_d$, $\mu_2=\mu_s$ and $T$. In the following we make the convention that $\langle\!\langle\bar{q}_1q_1\rangle\!\rangle=\langle\!\langle\bar{q}_uq_u\rangle\!\rangle=\langle\!\langle\bar{q}_dq_d\rangle\!\rangle$ and  $\langle\!\langle\bar{q}_2q_2\rangle\!\rangle=\langle\!\langle\bar{q}_sq_s\rangle\!\rangle$. Here the $\mathrm{SU(2)}$ flavor symmetry has been required. Let us specify the notations of these susceptibilities, taking the chiral susceptibilities as examples, $\chi^{ij}_s\equiv-\frac{\partial\langle\!\langle\bar{q}_iq_i\rangle\!\rangle}{\partial m_j}$ where $i,j\in\{1,2\}$. These susceptibilities are coupled with each other as follows:
\begin{widetext}
\begin{gather}
\begin{split}
 \hspace{0mm}\chi_s^{11}
 =&\chi_{s}^{1(0)}(1+4G_1\chi_s^{11}-2K\langle\!\langle\bar{q}_2q_2\rangle\!\rangle\chi_s^{11}-2K\langle\!\langle\bar{q}_1q_1\rangle\!\rangle\chi_s^{21})+\chi_{vs}^{1(0)}(4G_2\chi_m^{11}), \label{eq:chisc11}
\end{split}\\
\begin{split}
 \hspace{0mm}\chi_s^{21}
  =&\chi_{s}^{2(0)}(4G_1\chi_s^{21}-4K\langle\!\langle\bar{q}_1q_1\rangle\!\rangle\chi_s^{11})+\chi_{vs}^{2(0)}(4G_2\chi_m^{21}), \label{eq:chisc21}
\end{split}\\
\begin{split}
 \hspace{0mm}\chi_m^{11}
 =&\chi_{m}^{1(0)}(1+4G_1\chi_s^{11}-2K\langle\!\langle\bar{q}_2q_2\rangle\!\rangle\chi_s^{11}-2K\langle\!\langle\bar{q}_1q_1\rangle\!\rangle\chi_s^{21})+\chi_{q}^{1(0)}(4G_2\chi_m^{11}), \label{eq:chimc11}
\end{split}\\
\begin{split}
 \hspace{0mm}\chi_m^{21}
  =&\chi_{m}^{2(0)}(4G_1\chi_s^{21}-4K\langle\!\langle\bar{q}_1q_1\rangle\!\rangle\chi_s^{11})+\chi_{q}^{2(0)}(4G_2\chi_m^{21}).\label{eq:chimc21}
\end{split}
\end{gather}
\begin{gather}
\begin{split}
 \hspace{0mm}\chi_s^{22}
 =&\chi_{s}^{2(0)}(1+4G_1\chi_s^{22}-4K\langle\!\langle\bar{q}_1q_1\rangle\!\rangle\chi_s^{12})+\chi_{vs}^{2(0)}(4G_2\chi_m^{22}), \label{eq:chisc22}
\end{split}\\
\begin{split}
 \hspace{0mm}\chi_s^{12}
  =&\chi_{s}^{1(0)}(4G_1\chi_s^{12}-2K\langle\!\langle\bar{q}_1q_1\rangle\!\rangle\chi_s^{22}-2K\langle\!\langle\bar{q}_2q_2\rangle\!\rangle\chi_s^{12})+\chi_{vs}^{1(0)}(4G_2\chi_m^{12}), \label{eq:chisc12}
\end{split}\\
\begin{split}
 \hspace{0mm}\chi_m^{22}
 =&\chi_{m}^{2(0)}(1+4G_1\chi_s^{22}-4K\langle\!\langle\bar{q}_1q_1\rangle\!\rangle\chi_s^{12})+\chi_{q}^{2(0)}(4G_2\chi_m^{22}), \label{eq:chimc22}
\end{split}\\
\begin{split}
 \hspace{0mm}\chi_m^{12}
  =&\chi_{m}^{1(0)}(4G_1\chi_s^{12}-2K\langle\!\langle\bar{q}_1q_1\rangle\!\rangle\chi_s^{22}-2K\langle\!\langle\bar{q}_2q_2\rangle\!\rangle\chi_s^{12})+\chi_{q}^{1(0)}(4G_2\chi_m^{12}). \label{eq:chimc12}
\end{split}
\end{gather}
\begin{gather}
\begin{split}
 \hspace{0mm}\chi_{vs}^{11}
 =&\chi_{s}^{1(0)}(4G_1\chi_{vs}^{11}-2K\langle\!\langle\bar{q}_2q_2\rangle\!\rangle\chi_{vs}^{11}-2K\langle\!\langle\bar{q}_1q_1\rangle\!\rangle\chi_{vs}^{21})+\chi_{vs}^{1(0)}(1+4G_2\chi_q^{11}), \label{eq:chivsc11}
\end{split}\\
\begin{split}
 \hspace{0mm}\chi_{vs}^{21}
  =&\chi_{s}^{2(0)}(4G_1\chi_{vs}^{21}-4K\langle\!\langle\bar{q}_1q_1\rangle\!\rangle\chi_{vs}^{11})+\chi_{vs}^{2(0)}(4G_2\chi_q^{21}), \label{eq:chivsc21}
\end{split}\\
\begin{split}
 \hspace{0mm}\chi_q^{11}
 =&\chi_{m}^{1(0)}(4G_1\chi_{vs}^{11}-2K\langle\!\langle\bar{q}_2q_2\rangle\!\rangle\chi_{vs}^{11}-2K\langle\!\langle\bar{q}_1q_1\rangle\!\rangle\chi_{vs}^{21})+\chi_{q}^{1(0)}(1+4G_2\chi_q^{11}), \label{eq:chiqc11}
\end{split}\\
\begin{split}
 \hspace{0mm}\chi_q^{21}
  =&\chi_{m}^{2(0)}(4G_1\chi_{vs}^{21}-4K\langle\!\langle\bar{q}_1q_1\rangle\!\rangle\chi_{vs}^{11})+\chi_{q}^{2(0)}(4G_2\chi_q^{21}). \label{eq:chiqc21}
\end{split}
\end{gather}
\begin{gather}
\begin{split}
 \hspace{0mm}\chi_{vs}^{22}
 =&\chi_{s}^{2(0)}(4G_1\chi_{vs}^{22}-4K\langle\!\langle\bar{q}_1q_1\rangle\!\rangle\chi_{vs}^{12})+\chi_{vs}^{2(0)}(1+4G_2\chi_q^{22}), \label{eq:chivsc22}
\end{split}\\
\begin{split}
 \hspace{0mm}\chi_{vs}^{12}
  =&\chi_{s}^{1(0)}(4G_1\chi_{vs}^{12}-2K\langle\!\langle\bar{q}_1q_1\rangle\!\rangle\chi_{vs}^{22}-2K\langle\!\langle\bar{q}_2q_2\rangle\!\rangle\chi_{vs}^{12})+\chi_{vs}^{1(0)}(4G_2\chi_q^{12}), \label{eq:chivsc12}
\end{split}\\
\begin{split}
 \hspace{0mm}\chi_q^{22}
 =&\chi_{m}^{2(0)}(4G_1\chi_{vs}^{22}-4K\langle\!\langle\bar{q}_1q_1\rangle\!\rangle\chi_{vs}^{12})+\chi_{q}^{2(0)}(1+4G_2\chi_q^{22}), \label{eq:chiqc22}
\end{split}\\
\begin{split}
 \hspace{0mm}\chi_q^{12}
  =&\chi_{m}^{1(0)}(4G_1\chi_{vs}^{12}-2K\langle\!\langle\bar{q}_1q_1\rangle\!\rangle\chi_{vs}^{22}-2K\langle\!\langle\bar{q}_2q_2\rangle\!\rangle\chi_{vs}^{12})+\chi_{q}^{1(0)}(4G_2\chi_q^{12}). \label{eq:chiqc12}
\end{split}
\end{gather}
\begin{gather}
\begin{split}
 \hspace{-1mm}\chi_T^1
 \!\!=\!&\chi_{s}^{1(0)}\!(\!4G_1\chi_T^1\!\!-\!2K\langle\!\langle\bar{q}_2q_2\rangle\!\rangle\chi_T^1\!\!-\!2K\langle\!\langle\bar{q}_1q_1\rangle\!\rangle\chi_T^2\!)\!+\!\chi_{vs}^{1(0)}\!(\!4G_2\chi_n^1\!)\!+\! M_1\beta\chi_q^{1(0)}\!\!-\!\mu_{r1}\beta\chi_m^{1(0)}, \label{eq:chiT1}
\end{split}\\
\begin{split}
 \hspace{0mm}\chi_T^2
  =&\chi_{s}^{2(0)}(4G_1\chi_T^2-4K\langle\!\langle\bar{q}_1q_1\rangle\!\rangle\chi_T^1)+\chi_{vs}^{2(0)}(4G_2\chi_n^2)+M_2\beta\chi_q^{2(0)}-\mu_{r2}\beta\chi_m^{2(0)}, \label{eq:chiT2}
\end{split}\\
\begin{split}
 \hspace{-1mm}\chi_n^1
 =&\chi_{m}^{1(0)}(4G_1\chi_T^1\!-\!2K\langle\!\langle\bar{q}_2q_2\rangle\!\rangle\chi_T^1\!-\!2K\langle\!\langle\bar{q}_1q_1\rangle\!\rangle\chi_T^2)\!+\!\chi_{q}^{1(0)}(4G_2\chi_n^1)\!-\!\mu_{r1}\beta\chi_q^{1(0)}\!+\!\chi_t^{1(0)}, \label{eq:chin1}
\end{split}\\
\begin{split}
 \hspace{0mm}\chi_n^2
  =&\chi_{m}^{2(0)}(4G_1\chi_T^2-4K\langle\!\langle\bar{q}_1q_1\rangle\!\rangle\chi_T^1)+\chi_{q}^{2(0)}(4G_2\chi_n^2)-\mu_{r2}\beta\chi_q^{2(0)}+\chi_t^{2(0)}. \label{eq:chin2}
\end{split}
\end{gather}
\end{widetext}
where $\chi_T^i\equiv\frac{\partial\langle\!\langle\bar{q}_iq_i\rangle\!\rangle}{\partial T}$, $\chi_n^i\equiv\frac{\partial\langle\!\langle q_i^{\dagger}q^{}_i\rangle\!\rangle}{\partial T}$. The superscripts $^{1(0)}$, $^{2(0)}$ represent the free susceptibilities of the u(d)-quark sector and s-quark sector, respectively.
\begin{equation}
\chi_t^{(0)}=\frac{N_c}{\pi^2}\int_0^\Lambda p^2E\beta^2h(\mu){\ud}p.
\end{equation}
The inputs of variables of the free susceptibilities denoted with superscript $^{(0)}$ and functions $\chi_t^{(0)}$ mentioned in Eqs. (\ref{eq:chisc11})-(\ref{eq:chin2}) are constituent quark mass $M$ and renormalized chemical potential $\mu_r$ instead of current quark mass $m$ and bare chemical potential $\mu$.  Now five sets of linear equations for these susceptibilities are presented here. One can solve these sets of equations with the simple methods of linear algebra (set $\mu_1=\mu_2$ again), and then these susceptibilities will be at hand.
The same as in the SU(2) case~\cite{PhysRevD.88.114019}, we use the peaks of various susceptibilities $\chi_s^{11},\chi_{vs}^{11}$ and $\chi_T^{1}$ as the criteria to locate the critical crossover region. Fig. \ref{fig:phasedia} reveals an obvious critical band in the crossover region. Hence the same conclusion as in the SU(2) case can be drawn in the SU(3) case: it is more suitable to define a critical band rather than an exclusive line in the crossover region.

\section{Critical exponents for various susceptibilities}

As is well known that a second-order phase transition occurs at the CEP, where the correlation length tends to infinity and these susceptibilities diverge~\cite{costa2007qcd,PhysRevD.67.014028,costa2008thermodynamics,schaefer2007susceptibilities}. Here in the SU(3) case the CEP is located at $(T^{\mathrm{CEP}},\mu^{\mathrm{CEP}})=(67.73~\mathrm{MeV}, 318.42~\mathrm{MeV})$~\cite{costa2007qcd}  with $\langle\!\langle\bar{q}_uq_u\rangle\!\rangle^{1/3}=-191.9~\mathrm{MeV}$, $\langle\!\langle\bar{q}_sq_s\rangle\!\rangle^{1/3}=-251.8~\mathrm{MeV}$, $\langle\!\langle q^\dagger_u q^{}_u\rangle\!\rangle=3.171\times10^6~\mathrm{MeV}^3$ and $\langle\!\langle q^\dagger_s q^{}_s\rangle\!\rangle=2.424\times10^5~\mathrm{MeV}^3$, while in the SU(2) case, $(T^{\mathrm{CEP}},\mu^{\mathrm{CEP}})=(31.95~\mathrm{MeV}, 346.64~\mathrm{MeV})$  with $\langle\!\langle\bar{q}_uq_u\rangle\!\rangle^{1/3}=-206.5~\mathrm{MeV}$ and $\langle\!\langle q^\dagger_u q^{}_u\rangle\!\rangle=2.431\times10^6~\mathrm{MeV}^3$ using the parameter set and equations in the Ref.~\cite{PhysRevD.88.114019}. The critical behavior of the quantities of interest in the vicinity of the CEP can be well described by a set of indices, the so-called critical exponents. The motivation for this study arises from the fundamental and universal phase transition considerations. Some analysis based on nonperturbative renormalization group method makes very big yet not enough progress in the recent years~\cite{Bohr:2000gp,Schaefer:2004en,schaefer2007susceptibilities,Schaefer:2007pw,Aoki:1999dv,aoki2000analysis,aoki2000introduction,Aoki:2014ova,aoki2015rg}.

We adopt the quantity  $\Delta\langle\!\langle\bar{q}_iq_i\rangle\!\rangle=\langle\!\langle\bar{q}_iq_i\rangle\!\rangle_W-\langle\!\langle\bar{q}_iq_i\rangle\!\rangle_N$ or $\Delta\langle\!\langle q^\dagger_iq^{}_i\rangle\!\rangle=\langle\!\langle q^\dagger_iq^{}_i\rangle\!\rangle_W-\langle\!\langle q^\dagger_iq^{}_i\rangle\!\rangle_N$, along the first-order transition line towards the CEP, as the order parameter~\cite{PhysRevD.58.096007} by analogy with the liquid-gas phase transition, which falls into the $\text{Z}_2$ universality class, namely, the 3-dimensional Ising universality class~\cite{PhysRevC.67.025203,pelissetto2002critical}. The subscript $N$ and $W$ denote the Nambu phase (chiral symmetry broken) and the Wigner phase (chiral symmetry partially restored), respectively.

The determination of these critical exponents, which govern the strength of these susceptibilities' divergence, depends on whether the path to approach the CEP is asymptotically parallel to the first order transition line or not~\cite{PhysRevA.2.1047}. Indeed, the temperature $t$ and magnetic field $H$ of the 3D Ising model are mapped as
certain linear combinations of $T$ and $\mu$ on the $T-\mu$ plane near the CEP. To be specific, the $t$ direction is tangential to the first-order transition line while the $H$ direction is not. If the path is parallel to the $\mu$-axis in the $T-\mu$ plane, from lower (higher) $\mu$'s towards the critical $\mu^{\mathrm{CEP}}$, at fixed temperature $T=T^{\mathrm{CEP}}$ (marked by $\rightarrow$ ($\leftarrow$) in the following tables), the critical exponents $\epsilon_i^{}$ ($\epsilon_i^\prime$) are defined as follows:
\begin{gather}
\begin{split}
\ln\chi_i=-\epsilon_i\ln\mid\mu-\mu^{\mathrm{CEP}}\mid+c_i,
\end{split}\\
\begin{split}
\ln\chi_i=-\epsilon_i^\prime\ln\mid\mu-\mu^{\mathrm{CEP}}\mid+c_i^{}.
\end{split}
\end{gather}
where $c_i$ is a constant hereafter.

We can also choose the path parallel to the $T$-axis to approach the CEP and define the corresponding critical exponents following the same convention as above (the direction from lower $T$'s towards $T^{\mathrm{CEP}}$ is marked by $\uparrow$ and, conversely, $\downarrow$  ).

According to the definition of the baryon number susceptibility, in the SU(3) case,
\begin{equation}
\chi_B=\frac{1}{3}\sum_{i=u,d,s}(\frac{\partial\langle\!\langle q_i^\dagger q_i^{}\rangle\!\rangle}{\partial\mu_i})_T=\frac{2\chi_q^{11}+\chi_{q}^{22}}{3}.
\end{equation}
Since $\chi_q^{11}$ is at least 2 orders of magnitude larger than $\chi_q^{22}$ near the CEP in our calculation, the critical behavior of $\chi_B$ is mainly governed by $\chi_q^{11}$. While in the SU(2) case, $\chi_B=\chi_q^{11}$.

The critical exponents $\beta_i$  for the order parameters are defined as follows:
\begin{gather}
\begin{split}
\ln\Delta\langle\!\langle\bar{q}_iq_i\rangle\!\rangle=\frac{\bar{\beta_i}}{2}\ln\mid(T-T^{\mathrm{CEP}})^2+(\mu-\mu^{\mathrm{CEP}})^2\mid+c_i,
\end{split}\\
\begin{split}
\ln\Delta\langle\!\langle q^\dagger_iq^{}_i\rangle\!\rangle=\frac{\beta^\dagger_i}{2}\ln\mid (T-T^{\mathrm{CEP}})^2+(\mu-\mu^{\mathrm{CEP}})^2\mid+c_i,
\end{split}
\end{gather}
where the path to approach the CEP is along the first order transition line.

For comparison with the O(4) spin model universality class predictions, we also consider the critical behavior of the $\langle\!\langle\bar{q}_iq_i\rangle\!\rangle$ ($i=u,d$) in the chiral limit at vanishing chemical potential and from lower $T$'s towards the critical temperature $T_{\mathrm{c}}$ (located at 193.22 $\mathrm{MeV}$ and 191.80 $\mathrm{MeV}$ in SU(2) and SU(3) case, respectively) :
\begin{equation}
\ln\mid\langle\!\langle\bar{q}_iq_i\rangle\!\rangle\mid=\beta_i\ln\mid T-T_{\mathrm{c}}\mid+c_i.
\end{equation}
\begin{table*}[t]
\centering
	\caption{Critical exponents for order parameter}
	\begin{tabular}{|c|p{3.5cm}<{\centering}|p{2cm}<{\centering}|p{2cm}<{\centering}|c|}
		\hline
		Quantity                                             & Critical exponents   & SU(2) NJL      & SU(3) NJL & Universality \\
		\hline
        $\Delta\langle\!\langle\bar{q}_uq_u\rangle\!\rangle$     & $\bar\beta_u$        & 0.49$\pm$0.02  & 0.51$\pm$0.01    &1/2\\
		$\Delta\langle\!\langle\bar{q}_sq_s\rangle\!\rangle$     & $\bar\beta_s$        &                & 0.50$\pm$0.02    &1/2\\
		$\Delta\langle\!\langle q^\dagger_u q^{}_u\rangle\!\rangle$ & $\beta^\dagger_u$    & 0.48$\pm$0.02  & 0.51$\pm$0.02    &1/2\\
		$\Delta\langle\!\langle q^\dagger_s q^{}_s\rangle\!\rangle$ & $\beta^\dagger_s$    &                & 0.50$\pm$0.02    &1/2\\
		$\langle\!\langle\bar{q}_uq_u\rangle\!\rangle$           & $\beta_u$            & 0.50$\pm$0.01  & 0.49$\pm$0.01    &1/2\\
        \hline  	
\end{tabular}\label{4}
\end{table*}

\begin{table*}[tb]
\centering
\caption{Critical exponents  (parallel to $\mu$ axis)}
\begin{tabular}{|c|p{4.1cm}<{\centering}|p{2cm}<{\centering}|p{2cm}<{\centering}|}
\hline
Quantity & Critical exponents/Path                    & SU(2) NJL          & SU(3) NJL     \\
\hline
$\chi_q^{11}$ & $\epsilon_q^{11\,}/\rightarrow$       & 0.67$\pm$0.01         & 0.69$\pm$0.01   \\
         &  $\epsilon_q^{11\prime}/\leftarrow$        & 0.68$\pm$0.01          & 0.69$\pm$0.01   \\
$\chi_s^{11}$ & $\epsilon_s^{11\,}/\rightarrow$       & 0.69$\pm$0.01        &0.70$\pm$0.01       \\
         &  $\epsilon_s^{11\prime}/\leftarrow$        & 0.66$\pm$0.01       & 0.65$\pm$0.02      \\
$\chi_{vs}^{11}$ & $\epsilon_{vs}^{11\,}/\rightarrow$ & 0.68$\pm$0.01     & 0.71$\pm$0.01     \\
         &  $\epsilon^{11\prime}_{vs}/\leftarrow$     & 0.67$\pm$0.01     & 0.69$\pm$0.01   \\
$\chi_T^1$ & $\epsilon_T^1\,\,/\rightarrow$           & $0.65\pm$0.01        & 0.67$\pm$0.02                 \\
            & $\epsilon_T^{1\prime\,}/\leftarrow$     & 0.69$\pm$0.01    & 0.71$\pm$0.02                  \\
$\chi_n^1$ & $\epsilon_n^1\,\,/\rightarrow$           & 0.64$\pm$0.01     & 0.65$\pm$0.02                  \\
           & $\epsilon_n^{1\prime\,}/\leftarrow$      & 0.69$\pm$0.02    & 0.70$\pm$0.01                 \\
\hline
\end{tabular}\label{1}
\end{table*}

\begin{table*}[!htb]
\centering
	\caption{Critical exponents (parallel to $T$ axis)}
	\begin{tabular}{|c|p{4.1cm}<{\centering}|p{2cm}<{\centering}|p{2cm}<{\centering}|}
		\hline
		Quantity & Critical exponents/Path              & SU(2) NJL              & SU(3) NJL    \\
		\hline
        $\chi_q^{11}$ & $\epsilon_q^{11\,}/\uparrow$      & 0.68$\pm$0.01          & 0.70$\pm$0.01   \\
		&  $\epsilon_q^{11\prime}/\downarrow $          & 0.67$\pm$0.01          &  0.68$\pm$0.01\\
		$\chi_s^{11}$ & $\epsilon_s^{11\,}/\uparrow$      & 0.69$\pm$0.01          & 0.71$\pm$0.01   \\
		&  $\epsilon_s^{11\prime}/\downarrow$           & 0.66$\pm$0.01          & 0.64$\pm$0.01    \\
		$\chi_{vs}^{11}$ & $\epsilon_{vs}^{11\,}/\uparrow$& 0.68$\pm$0.01          & 0.70$\pm$0.01\\
		&  $\epsilon^{11\prime}_{vs}/\downarrow$        & 0.67$\pm$0.01          & 0.68$\pm$0.01   \\
		$\chi_T^1$ & $\epsilon_T^1\,\,/\uparrow$            & 0.66$\pm$0.01          & 0.68$\pm$0.01\\
		& $\epsilon_T^{1\prime\,\,}/\downarrow$             & 0.68$\pm$0.01          & 0.70$\pm$0.01     \\
		$\chi_n^1$ & $\epsilon_n^1\,\,/\uparrow$            & 0.66$\pm$0.01          & 0.66$\pm$0.02\\
		& $\epsilon_n^{1\prime\,\,}/\downarrow$             & 0.69$\pm$0.01          & 0.69$\pm$0.01\\
	    \hline
	\end{tabular} \label{2}
\end{table*}

\begin{table*}[!htb]
\centering
	\caption{Critical exponents (for s quark sector)}
	\begin{tabular}{|c|p{3.1cm}<{\centering}|p{3.5cm}<{\centering}|p{3.5cm}<{\centering}|}
		\hline
		Quantity & Critical exponents          & SU(3) NJL($\parallel \mu$ axis)      & SU(3) NJL($\parallel T$ axis)  \\
		\hline
        $\chi_q^{22}$ & $\epsilon_q^{22}$      & 0.68$\pm$0.04/$\rightarrow$          & 0.68$\pm$0.04/$\uparrow$\\
		&  $\epsilon_q^{22\prime}$             & 0.65$\pm$0.02/$\leftarrow$           & 0.65$\pm$0.02/$\downarrow$\\
		$\chi_s^{22}$ & $\epsilon_s^{22}$      & 0.67$\pm$0.03/$\rightarrow$          & 0.67$\pm$0.03/$\uparrow$\\
		&  $\epsilon_s^{22\prime}$             & 0.66$\pm$0.02/$\leftarrow$           & 0.65$\pm$0.02/$\downarrow$\\
		$\chi_{vs}^{22}$ & $\epsilon_{vs}^{22}$& 0.67$\pm$0.03/$\rightarrow$          & 0.68$\pm$0.03/$\uparrow$\\
		&  $\epsilon^{22\prime}_{vs}$          & 0.65$\pm$0.02/$\leftarrow$           & 0.65$\pm$0.02/$\downarrow$\\
		$\chi_T^2$ & $\epsilon_T^2\,\,$            & 0.69$\pm$0.02/$\rightarrow$          & 0.70$\pm$0.02/$\uparrow$\\
		& $\epsilon_T^{2\prime\,\,}$               & 0.70$\pm$0.01/$\leftarrow$           & 0.70$\pm$0.01/$\downarrow$\\
		$\chi_n^2$ & $\epsilon_n^2\,\,$            & 0.68$\pm$0.02/$\rightarrow$          & 0.69$\pm$0.03/$\uparrow$\\
		& $\epsilon_n^{2\prime\,\,}$               & 0.68$\pm$0.01/$\leftarrow$           & 0.67$\pm$0.02/$\downarrow$\\
      \hline	
\end{tabular}\label{3}
\end{table*}

\begin{figure}[tbp]
\centering
\includegraphics[width=3in]{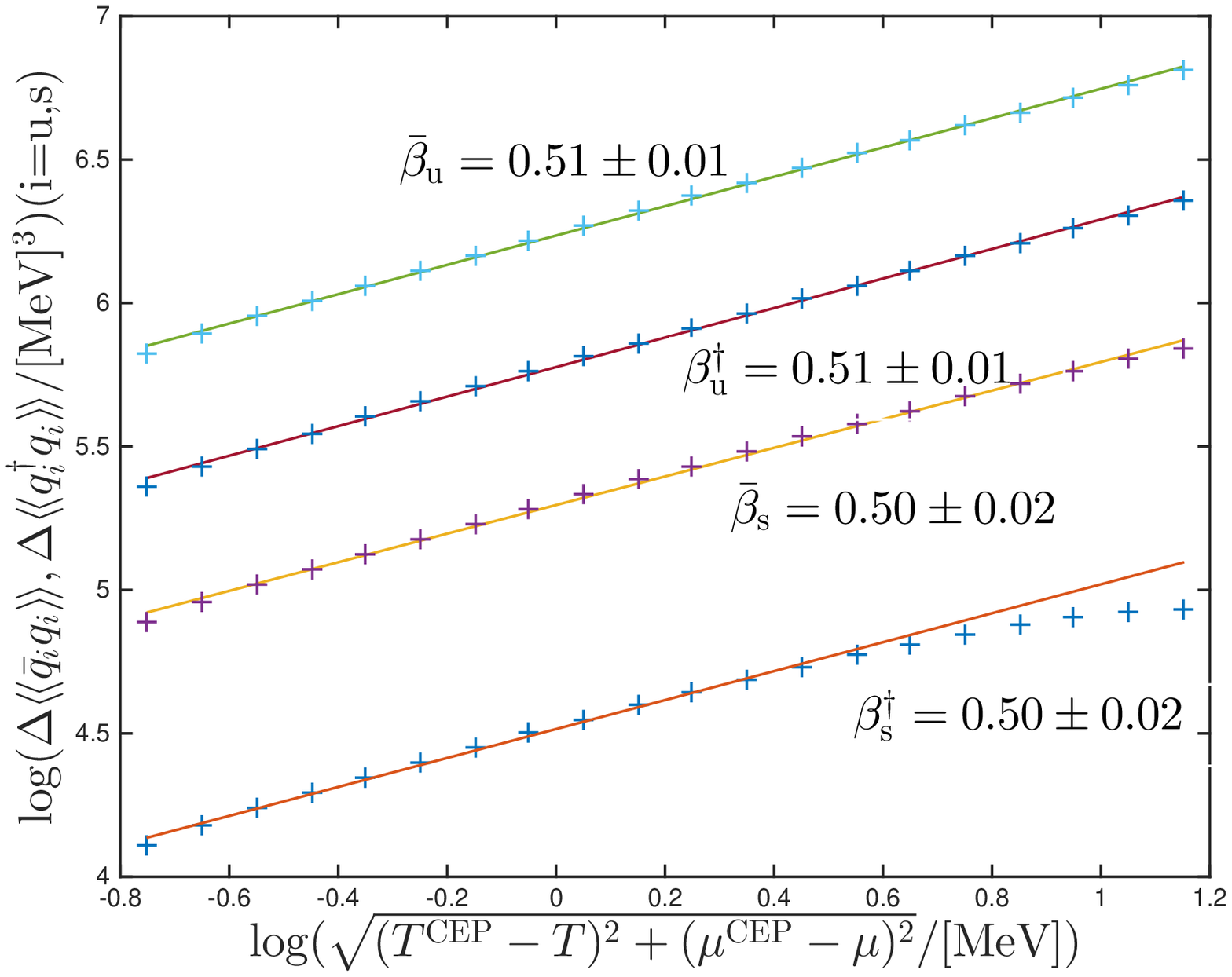}
\caption{(color online): Order parameters $\Delta\langle\!\langle\bar{q}_iq_i\rangle\!\rangle$, $\Delta\langle\!\langle q_i^\dagger q^{}_i\rangle\!\rangle$ ($i=u,s$) as a function of $\sqrt{(T^{\mathrm{CEP}}-T)^2+(\mu^{\mathrm{CEP}}-\mu)^2}$ along the first order transition line towards the CEP in the SU(3) case.}
\label{orderpara}
\includegraphics[width=3in]{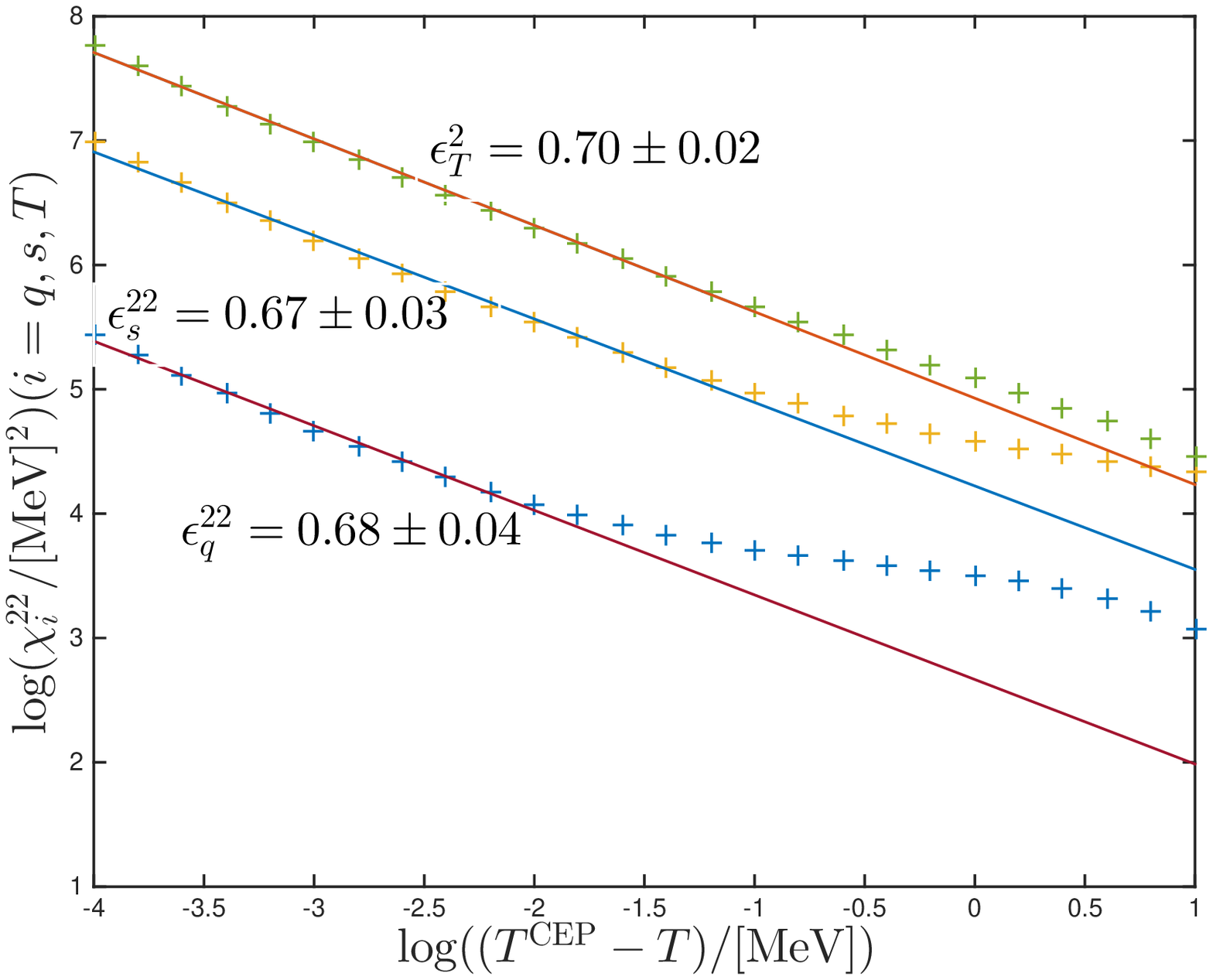}
\caption{(color online): Susceptibilities for the s-quark sector as a function of $T^{\mathrm{CEP}}-T$ at $\mu^{\mathrm{CEP}}$ in the SU(3) case.}
\label{lowerTsquark}
\includegraphics[width=3in]{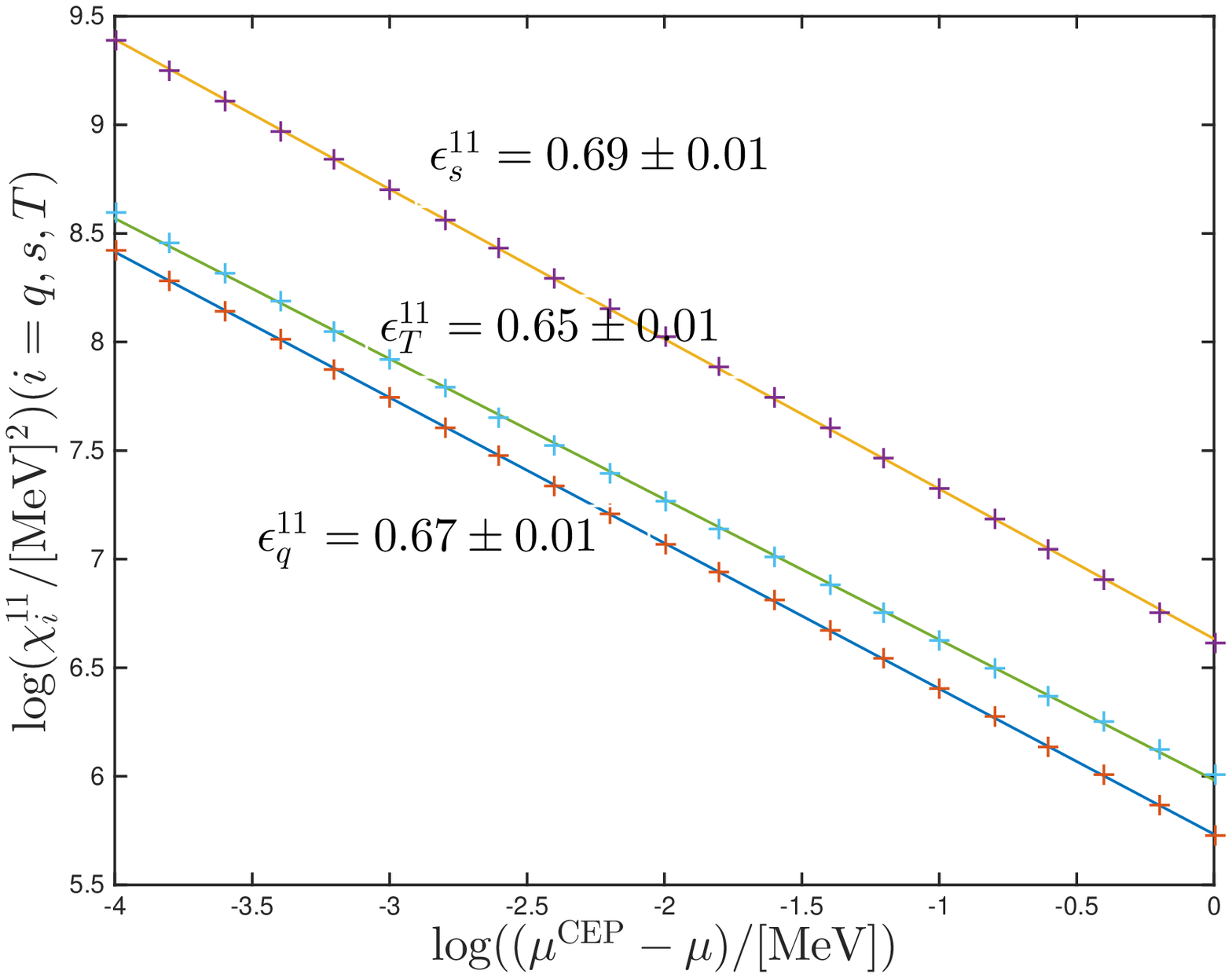}
\caption{(color online): Susceptibilities for the u(d)-quark sector as a function of $\mu^{\mathrm{CEP}}-\mu$ at $T^{\mathrm{CEP}}$ in the SU(2) case.}
\label{lowermuSU(2)}
\end{figure}

The critical exponents for the order parameters both in SU(2) and SU(3) cases are presented in Table \ref{4}. The $\beta$ values of $\Delta\langle\!\langle\bar{q}_iq_i\rangle\!\rangle$ and $\Delta\langle\!\langle q^\dagger_iq^{}_i\rangle\!\rangle$ towards the CEP along the first-order phase transition line are consistent with the theoretical estimated $\beta$ value $\frac{1}{2}$ in mean field theory instead of the $\beta$ value $0.30-0.34$ for the 3D Ising model ~\cite{pelissetto2002critical}. Logarithmic fitting procedures are shown in Fig. \ref{orderpara} taking the SU(3) case as an example. The $\beta$ value of $\langle\!\langle\bar{q}_uq_u\rangle\!\rangle$ at the critical temperature $T_{\mathrm{c}}$ and the vanishing chemical potential in the chiral limit is 0.50 (0.49) $\pm$ 0.01 in the SU(2) (SU(3)) case, which is also consistent with the predicted value $\frac{1}{2}$ in mean field theory instead of the O(4) spin model. These results are reasonable intuitively since the mean field approximation has been applied to the NJL model in this work.

The critical exponents of the various susceptibilities in the SU(3) case for the u(d)-quark sector are presented in Table \ref{1}, \ref{2} and those for the s-quark sector in Table \ref{3} and partly in Fig. \ref{lowerTsquark}. The critical exponents in the SU(2) case are also presented in Table \ref{1}, \ref{2} and partly in Fig. \ref{lowermuSU(2)}. It can be seen from Table \ref{1}-\ref{3} that these various susceptibilities share almost the same critical behavior near the CEP, that is to say, the $\epsilon$ values of these quantities are very close to the mean-field exponent $\frac{2}{3}$ no matter whether the path is parallel to the $T$-axis or $\mu$-axis and what the direction is chosen to approach the CEP both in SU(2) and SU(3) cases. It is reasonable since the critical fluctuation mixes in the quark number density fluctuation and chiral condensate fluctuation as well as in the thermal fluctuation at the CEP, due to finite current quark mass $m$ and quark chemical potential $\mu$~\cite{PhysRevD.67.014028,karsch2000three}. This mixture can be well understood from Eqs. (\ref{eq:chisc11})-(\ref{eq:chin2}) by the coupling relations of $\chi_q$, $\chi_s$ and $\chi_T$ (corresponding to the quark number density, chiral condensate and the thermal fluctuation, respectively), especially between different quark flavors. In Ref.~\cite{schaefer2012qcd} the path dependencies of these critical exponents near the CEP are also discussed in (Polyakov-)Quark-Meson model, which both confirm our result, and the interesting scaling crossing phenomenon in a narrow regime near CEP is demonstrated there. It's worth pointing out that the critical exponents of the baryon number susceptibility obtained in Ref.~\cite{costa2008thermodynamics} are reproduced here and they agree with the value of the mean field theory prediction $\frac{2}{3}$~\cite{PhysRevD.67.014028}. In addition, Fig. \ref{lowerTsquark} shows a crossover of different universality classes from the changes of the exponents related with the s-quark sector, which is consistent with the spirit of Ref.~\cite{PhysRevD.67.014028}. Therefore, it can be indicated that s-quark may has a smaller critical region for the CEP than u/d-quark does, from the comparison with the exponents related with the u(d)-quark sector presented in Fig. \ref{lowermuSU(2)}.

\section{Summary and conclusions}
In this paper we have strictly derived several sets of coupled equations for the various susceptibilities at finite temperature and quark chemical potential in the mean field approximation of (2+1)-flavor Nambu--Jona-Lasinio model for the first time. Then the rationality of using these susceptibilities as the criteria to determine the crossover region in the SU(3) case is discussed. It is thought to be more suitable to define a critical band rather than an exclusive line in the crossover region.

We have located the CEP at $(T^{\mathrm{CEP}},\mu^{\mathrm{CEP}})=(67.73~\mathrm{MeV}, 318.42~\mathrm{MeV})$ for $N_f=3$ and $(T^{\mathrm{CEP}},\mu^{\mathrm{CEP}})=(31.95~\mathrm{MeV}, 346.64~\mathrm{MeV})$ for $N_f=2$. The various susceptibilities and their corresponding critical exponents have been calculated in the SU(2) and SU(3) cases, respectively. For the first time our calculations roughly verify the mean-field prediction that near the CEP these susceptibilities share the same critical behavior no matter whether the path is parallel to the $T$-axis or $\mu$-axis and what the direction is chosen to approach the CEP, which is due to the mixture of the quark number density fluctuation and chiral condensate fluctuation as well as the thermal fluctuation at CEP in the presence of finite current quark mass $m$ and quark chemical potential $\mu$. The critical exponents of the order parameters towards the CEP along the first order transition line and towards the $T_c$ at vanishing quark chemical potential in the chiral limit both agree with the mean-field predictions. More insightful work beyond mean field approximation calls for more applications of the renormalization group method to this issue within the NJL model or other nonperturbative effective theories.

\acknowledgments
We would like to express our appreciation for helpful discussions with Yong-Hui Xia. This work is supported in part by the National Natural Science Foundation of China (under Grants No. 11275097, No. 11475085, and No. 11535005), and the Jiangsu Planned Projects for Postdoctoral Research Funds (under Grant No. 1402006C).



\bibliographystyle{apsrev4-1}
\bibliography{duyl}
\end{document}